\documentclass[10pt]{article}

\usepackage{amsmath}
\usepackage{amssymb}

\usepackage{cite}

\usepackage{gensymb}

\usepackage{graphicx}

\usepackage{microtype}
\DisableLigatures[f]{encoding = *, family = * }

\usepackage{setspace} 
\usepackage[labelfont=bf,labelsep=period,justification=raggedright]{caption}

\makeatletter
\renewcommand{\@biblabel}[1]{\quad#1.}
\makeatother

\date{}

\begin{document}

\begin{flushleft}
{\Large
\textbf{Global evidence for non-random dynamics in fish recruitment}
}
\\
Charles T. Perretti$^{1,3\ast}$, 
Stephan B. Munch$^2$,
Michael J. Fogarty$^3$
George Sugihara$^1$
\\
\bf{1} Scripps Institution of Oceanography, UCSD, La Jolla, CA 92093, USA
\\
\bf{2} National Marine Fisheries Service, National Oceanic and Atmospheric Administration,Santa Cruz, CA 95060, USA
\\
\bf{3} National Marine Fisheries Service, National Oceanic and Atmospheric Administration, Woods Hole, MA 02543, USA
\\
$\ast$E-mail: Charles.Perretti@noaa.gov
\end{flushleft}

\section*{Abstract}
Understanding what controls apparently random fluctuations in fish recruitment is a major challenge in fisheries science. Our current inability to anticipate recruitment failures has led to costly management actions and in some cases complete fishery collapse \cite{radovich_collapse_1982, whitmore_assessment_2013}. Time series observations of fish recruitment reflect an interplay of underlying population processes, environmental forcing, measurement error, and process error. Given that the error component is often very strong, an important unresolved question is whether any non-random signal can be uncovered in the annual fluctuations of recruitment time series. Here, we address this fundamental question in an analysis of 569 fish populations from a global database of recruitment. Using a nonparametric time series analysis method, we find overwhelming evidence for non-random dynamics operating in the recruitment process. Unlike previous explorations of this topic, our approach does not require the specification of a stock-recruitment model, and it can be used on recruitment estimates derived from a wide-range of abundance estimation methods. The evidence for non-randomness is robust across a wide range of fish families and abundance estimation methods. We also find that the statistical support for non-randomness increases consistently with the number of observations of a fish stock. This result provides the encouraging news that with continued observations, and the appropriate covariates, we should ultimately be able to uncover the mechanistic drivers of fish recruitment.

\section*{Introduction}
Uncovering the drivers of fluctuations in animal abundance is a fundamental goal of population biology. It is especially important in fisheries science where unexpected population failures may lead to large economic losses \cite{radovich_collapse_1982, whitmore_assessment_2013}. A major source of variability in most fish populations is in the recruitment stage, however, attempts to predict recruitment have met limited success. One explanation for this failure of prediction is that strong process error and measurement error effectively obscures any predictable signal. As a result, one might conclude that year-to-year fluctuations in recruitment are best viewed as functionally random.

The idea that recruitment is driven by stochastic events is not new. The role of the environment on larval survival was studied extensively in the early 20th century \cite{hjort_fluctuations_1914, hjort_fluctuations_1926, cushing_natural_1974, cushing_plankton_1990, lasker_relations_1978, lasker_factors_1981, sinclair_marine_1988}. Larval vulnerability was linked to growth which in turn was linked to stochastic environmental conditions \cite{houde_fish_1987, francis_does_1993}. The general conclusion of which was that small variations in survival due to environmental conditions can lead to large variations in future abundance. This, combined with demographic stochasticity which may be an important source of year-to-year variability in depleted populations \cite{fogarty_recruitment_1993}, may explain the commonly-observed high variation in recruitment time series. 

A counter to the random view was established in now-classic work by May and colleagues which showed that complex, apparently random, dynamics can arise in simple nonlinear systems \cite{may_simple_1976}. Therefore, an alternative explanation for recruitment variability is that it arises from intrinsic population dynamics coupled with environmental forcing. This should result in recruitment time series that contain a detectable non-random signal, albeit one that is convolved with stochasticity and observation error. 

Many attempts have been made to extract meaningful patterns from recruitment data. A common approach is to identify a covariate which might explain some of the variability in average recruitment. The spawner-recruit relationship is a well-known example of this that has met various levels of success \cite{iles_review_1994, myers_is_1996}. Changes in predation have also been identified as important (\cite{levin_interactive_1997}), as well as more recent studies which have highlighted the importance of environmental regimes \cite{szuwalski_examining_2014}. These studies provide firm evidence that recruitment is, in general, not random. However, they do not address our question of whether a non-random signal can be detected in the \textit{year-to-year} variations in \textit{univariate} time series of recruitment. This distinction is important, because even though a deterministic relationship may exist between recruitment and some covariate, this does not automatically mean that a non-random signal will be found in the resulting time series, especially when strong stochasticity and observation error are present.

Here we address this fundamental question by determining whether an underlying non-random signal can be detected in the annual fluctuations of recruit abundance in 569 time series of the Ransom A. Myers global stock-recruitment database (RAM, \cite{myers_http://www.mscs.dal.ca/myers/welcome.html_2014}). To do so, we use a Monte Carlo, nonparametric time series method which, unlike past studies based on parametric models, side-steps the problem of model-misspecification. We measure the strength of the non-random signal using a randomization test which estimates the probability that a population's recruitment time series originates from a random process ($P_{R}$, Fig.\ 1). Many of the time series in the RAM database are short and noisy, therefore a signal of non-randomness is noteworthy, if it exists.

\section*{Methods}

A recruit is defined as an individual that has recently reached a particular life-stage \cite{jennings_marine_2009}. In fisheries science, recruits are typically those that recently reached the minimum size to be legally harvested. The RAM database is a global database of time series records of recruit abundance encompassing over 600 fish populations and over 100 species from marine and freshwater environments (\cite{myers_http://www.mscs.dal.ca/myers/welcome.html_2014}, Fig. S2, Fig. S3). Time series lengths in the RAM database vary from as little as five years to as many as 74 years, with a mean length of 20.7 years. All populations with at least 10 years of data were included in the analysis, resulting in 569 populations.

Leave-one-out cross-validation was performed using Simplex projection \cite{sugihara_nonlinear_1990}, a nonparametric time series analysis method with one free parameter (embedding dimension, \textit{E}). Like all time series methods, it operates under the assumption that if the data-generating process is non-random, then the present state of the population should provide some information about the future state. Following Takens' theorem \cite{takens_detecting_1981}, Simplex predictions are made by finding similar past states and projecting them forward. Lags of the original time series are used to form a reconstruction of the high-dimensional ``attractor'', which is then used to make predictions (Fig. S4). The dimensionality of the reconstruction is determined by the embedding dimension (\textit{E}) which we restrict to values of one to four (Fig.\ S5). The advantage of this approach is that it is robust to model misspecification \cite{perretti_nonparametric_2012}, the validity of the approach is determined using a Monte Carlo test. 

Since the method used here is univariate, we are able to test the dynamics of fish populations that lack spawner abundance data as only recruit data is required. This expands the number of fish populations that we can investigate compared to previous studies that required both spawner and recruit data.

If recruit abundance is dominated by random forcing, present abundance will fail to predict future abundance. In this regard, we use a Monte Carlo test to determine whether the cross-validation fit for the real time series is significantly better than would be obtained from a random time series with the same statistical properties (i.e., the same empirical probability distribution and autoregressive properties). For each population, we fit an autoregressive model of order zero or one (chosen using AICc), which is then used to generate 1000 random time series (similar to the method of ``surrogate data,'' \cite{theiler_testing_1992}). The probability that the real time series is random ($P_{R}$) is obtained by comparing the Simplex fit for the real time series ($\rho_{real}$, Fig. 1a) to the distribution of Simplex fits for 1000 random time series ($\rho_{random}$, Fig.\ 1c). Cross-validation fit was measured as the correlation between predicted and observed recruit abundances. We allow for non-normal random time series by using an empirical density function when the best-fit autoregressive model is of order zero. 

If the recruitment time series are non-random, we expect to gain confidence in this with additional time series observations (previously described as ``convergence'' \cite{sugihara_detecting_2012}). In contrast, if the system is functionally stochastic, our confidence should decrease with additional observations. If convergence is occurring it will be represented by a decrease in probability random ($P_{R}$) with increasing time series length. We test for convergence by performing a beta-regression of time-series length on $P_{R}$. That is, we evaluate the relationship between the number of recruitment observations in a time series and the $P_{R}$ for that time series. If $P_{R}$ decreases as time series length increases we conclude this is evidence of convergence.

Importantly, probability random ($P_{R}$) is not expected to decrease merely due to an increase in time series length. To illustrate this point, we perform the Monte Carlo procedure on three example random datasets (Fig.\ 2a-c, e-g). For all datasets, the real time series are replaced with purely random time series of the same length and gaps (missing values) as the real time series. Three random datasets were simulated: 1) a standard-normal random variable (white noise), 2) an autocorrelated random variable with mean and return rate of the real time series (autocorrelated noise), and 3) a standard log-normal random variable (log-normal noise). All of the example random datasets fail to exhibit convergence.

\section*{Results and Discussion}

Importantly, none of the example random datasets showed evidence of convergence toward non-random dynamics, as the histograms of $P_R$ are not skewed towards low values (Fig.\ 2 a-c), nor is there convergence toward low $P_R$ with increasing time series length (Fig.\ 2 e-g).

In contrast to the example random datasets, the RAM database exhibited strong evidence for non-randomness. The most commonly observed value of $P_R$ for the RAM database was less than 0.03 (i.e., the mode of the distribution, Fig.\ 2d). Furthermore, the overall distribution of $P_{R}$ is significantly skewed towards low values ($p<1e^{-15}$, Kolmogorov-Smirnov test). In contrast, the three simulated random datasets had distributions of $P_{R}$ that were not significantly different from a uniform distribution ($p>0.8$, Fig.\ 2 a,b,c).

If recruitment time series have a detectable non-random signal, $P_{R}$ should diminish with additional observations (i.e., convergence should occur). This is observed in the RAM database. Populations with long time series have a lower $P_R$ on average than those with short time series (Fig.\ 2h). This trend was strongly significant (beta regression, $p<1e^{-15}$). Moreover, not only does $P_{R}$ decline, but the amount of variance explained by the nonparametric model increases with increasing time series length (Fig.\ 3). This suggests that with increasing observations the underlying dynamics become increasingly clear.

Despite the wide variety of methods used to estimate abundance in the RAM database, we observe convergence towards non-random dynamics in all major abundance estimation methods (those with $>$30 populations, Fig.\ 4, Fig.\ S6). This was also observed in all but one major family (Fig.\ 5, Fig.\ S7). Percidae is the only major family that fails to exhibit convergence (Fig.\ 5c), and is also the only major family that does not contain at least one long time series ($>$35 years). The rate of convergence (i.e., the slope of the beta-regression line) is proportional to the length of the longest time series (Fig.\ 5e). Thus, families with long time series are more likely to exhibit non-random dynamics, and we expect Percidae to exhibit non-random dynamics with additional observations. We also find that populations with significant high-order partial autocorrelation are more likely to be classified as non-random (Fig.\ S8), which again suggests that a non-random pattern exists in those time series.

In summary, using data only on recruit abundance, we uncover a widespread pattern of non-random dynamics in fish recruitment. This result is robust across all major abundance estimation methods and fish families. This builds upon previous studies which have identified covariates related to average levels of recruitment, as well as time series studies on smaller datasets (e.g., \cite{vert-pre_frequency_2013}). In contrast to the recruitment database, we find all of the simulated random datasets and the environmental time series fail to show evidence of non-random dynamics.

Our results suggest that with time we should be able to uncover the drivers of inter-annual recruitment fluctuations, however it is difficult to draw broad conclusions about mechanisms from this study alone. One possibility is that the patterns of recruitment in the RAM database are driven by persistent shifts in regimes which are driven by environmental forcing, or changes in predator and prey abundance \cite{minto_productivity_2013, szuwalski_examining_2014}. Multiple factors unique to each species are likely involved, and focused studies are required to uncover mechanisms. In this regard, mechanistic studies could be complimented by nonparametric methods to ensure robust conclusions.

Fish recruitment is a complex process operating on a range of temporal and spatial scales, therefore it is unsurprising that it is difficult to anticipate recruitment fluctuations. However here, using a simple time series approach, we have shown that non-random dynamics are widespread in recruitment. This suggests that prediction of fish recruitment should eventually be possible, as is already true in some cases (e.g, \cite{martino_recruitment_2010}). The increasing probability of observing non-random dynamics with additional observations suggests that predictability will also increase with continued observations. With time series data for key parallel variables it should be possible to identify the mechanisms that drive recruitment fluctuations. This provides the encouraging news that further work will allow improved forecasting and management of our wild fisheries.


\bibliographystyle{plos2009}

%
%
%

\section*{Figures}

%
\begin{figure}[htb]
\includegraphics[width=\textwidth]{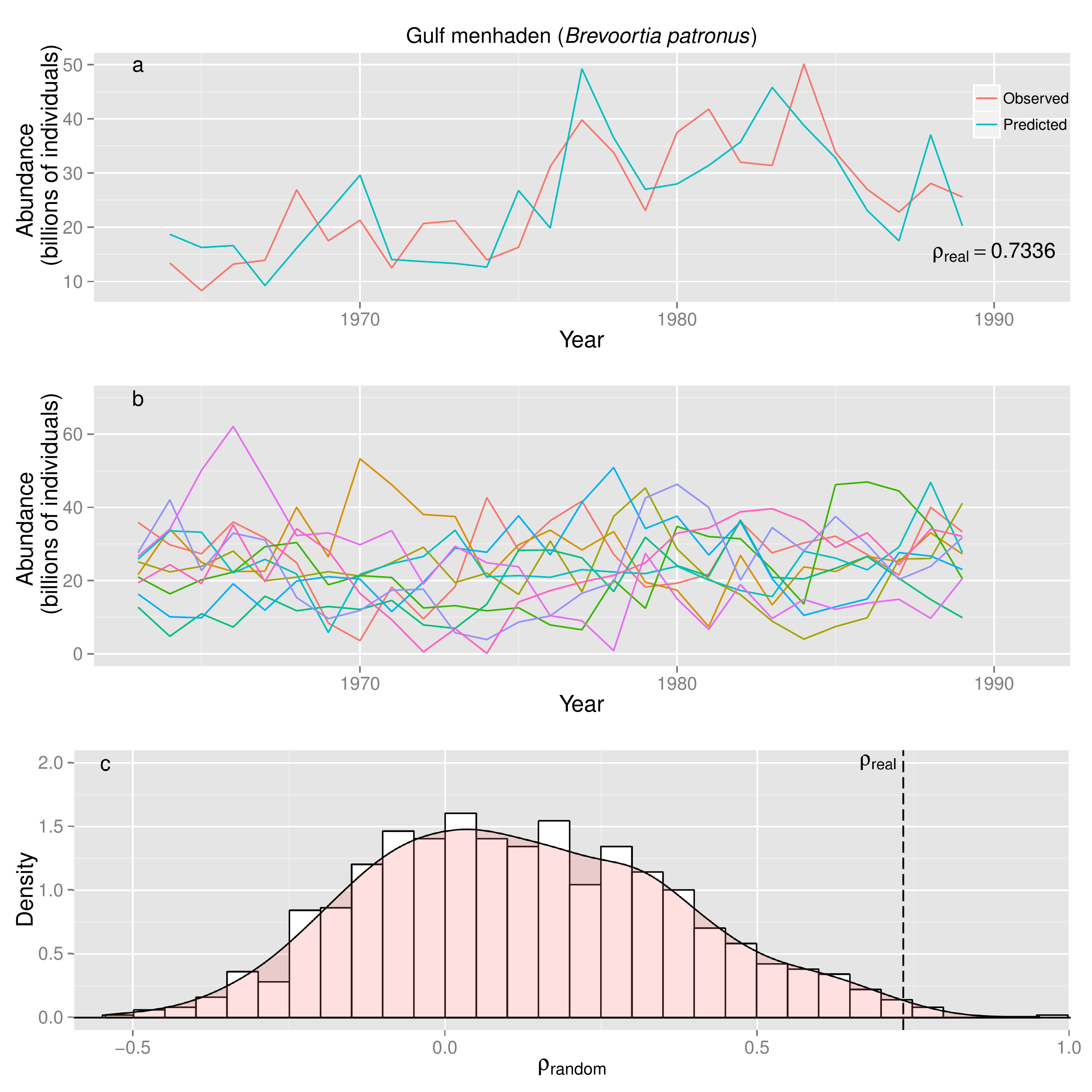}
\caption{{\bf Demonstration of the procedure used to measure the probability that a time series is random ($P_{R}$).} Predicted and observed time series for Gulf menhaden (\textit{Brevoortia patronus}) (a). Ten time series generated by the best-fit random model (autoregressive model with return rate = 0.6) (b). Histogram of correlations of predicted and observed for the random time series ($\rho_{random}$), and real time series ($\rho_{real}$) (c). $P_{R}$ is the area under the curve to the right of the dashed line in (c).}
\end{figure}

\begin{figure}[htb]
\includegraphics[width=\textwidth]{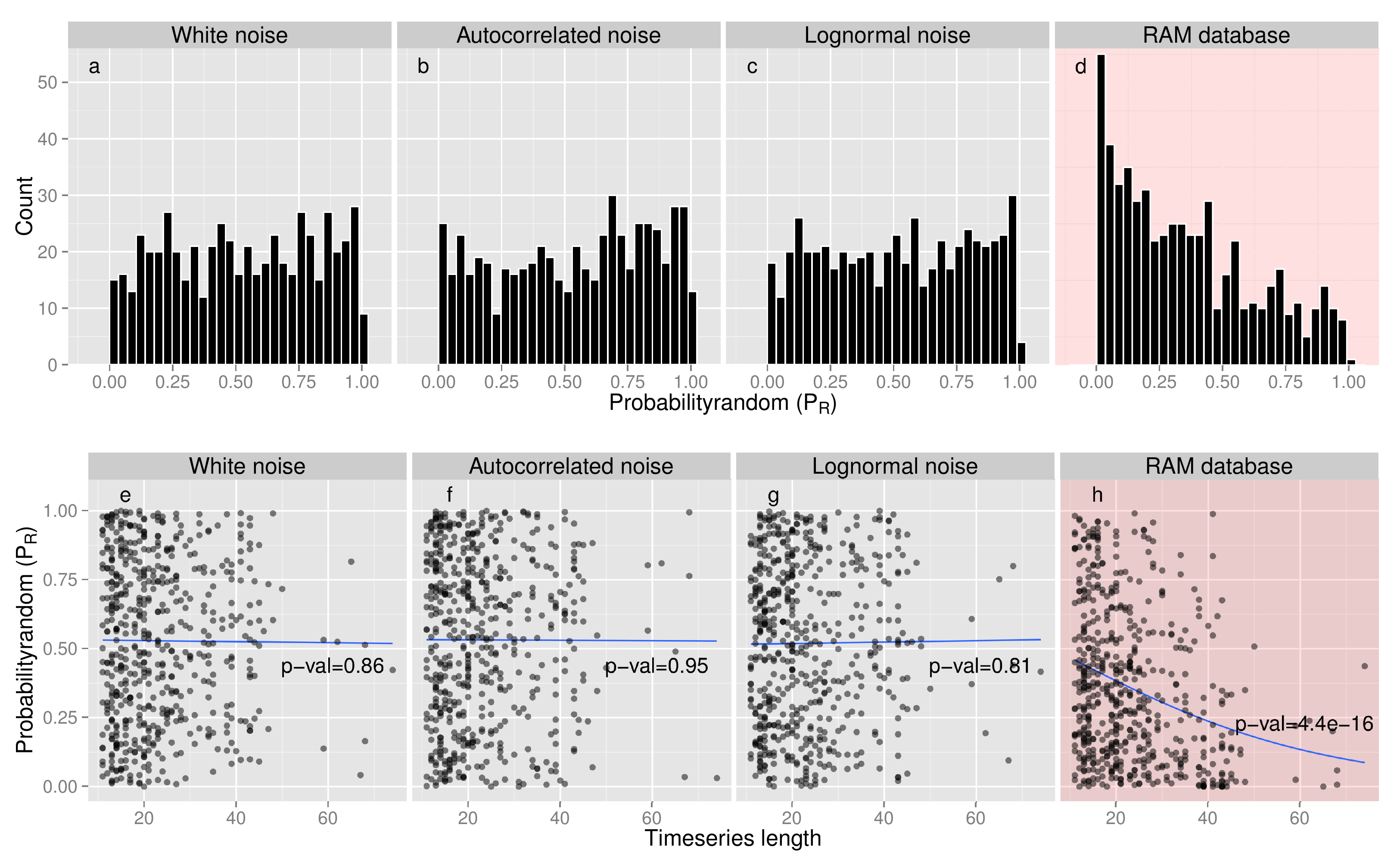}
\caption{{\bf Comparison of example random datasets and the RAM database.} Time series for the example random datasets were the same length and contained the same gaps (missing values) as the RAM database. Example random datasets are white noise (a,e), autocorrelated noise (b,f) and lognormal noise (c,g). Unlike the example datasets, the distribution for the RAM database is skewed towards low probability random (d), and exhibits a significant trend towards low probability random with increasing time series length (h).}
\end{figure}

\begin{figure}[htb]
\includegraphics[width=\textwidth]{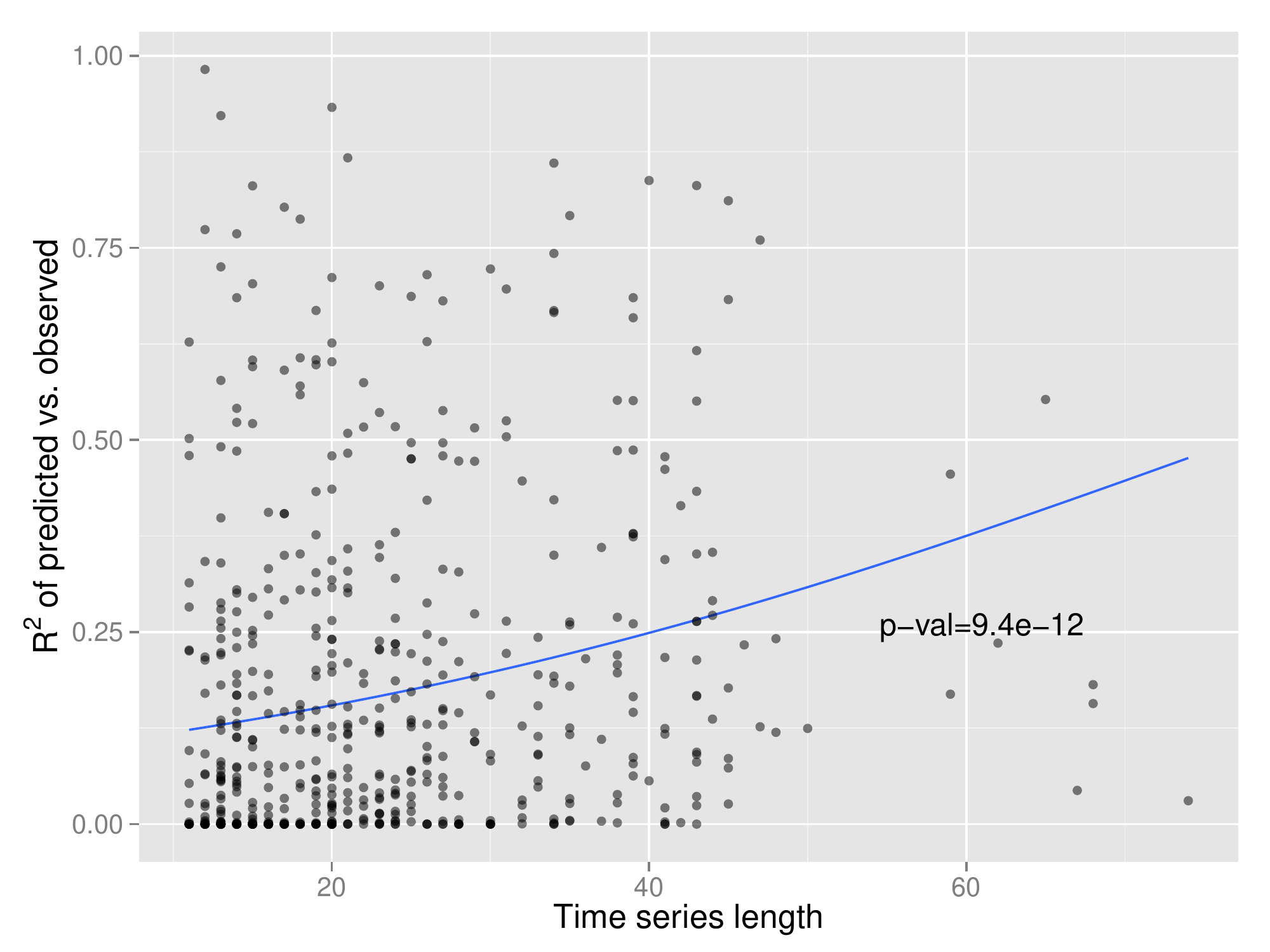}
\caption{{\bf As time series length increases, so does the R$^2$ of the predicted vs.\ observed for the nonparametric model fit.} Similar to the relationship between P$_R$ and time series length (Fig. 2h), this suggests that with increasing observations we are better able to capture the underlying dynamics of recruitment.}
\end{figure}

\begin{figure}[htb]
\includegraphics[width=\textwidth]{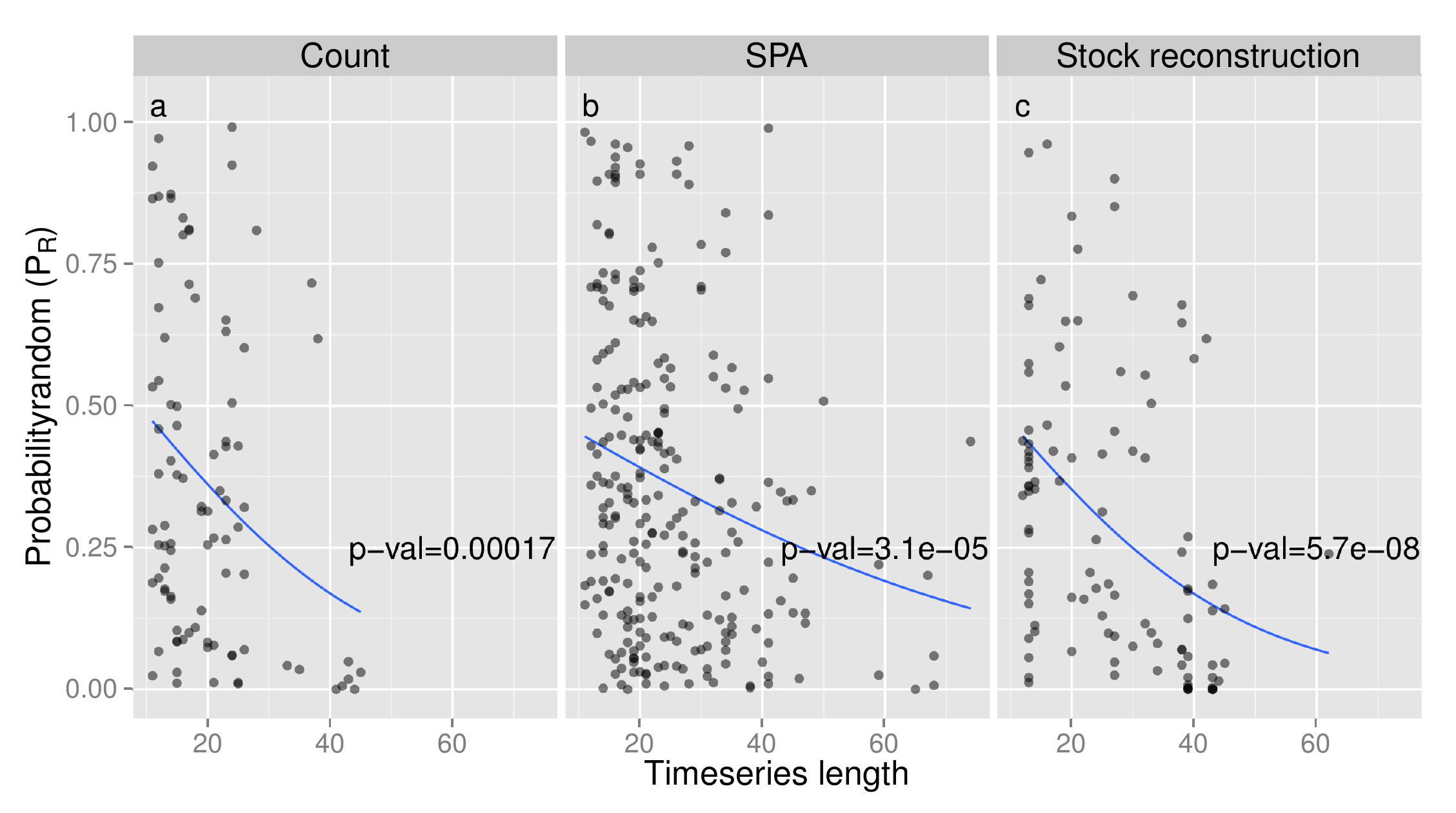}
\caption{{\bf Convergence across all major abundance estimation methods.} Abundance estimation methods with at least 30 populations are shown, they are Count (a), Sequential population analysis (SPA) (b), and Stock reconstruction (c). For all methods, probability random ($P_R$) declines significantly with time series length.}
\end{figure}

\begin{figure}[htb]
\includegraphics[width=\textwidth]{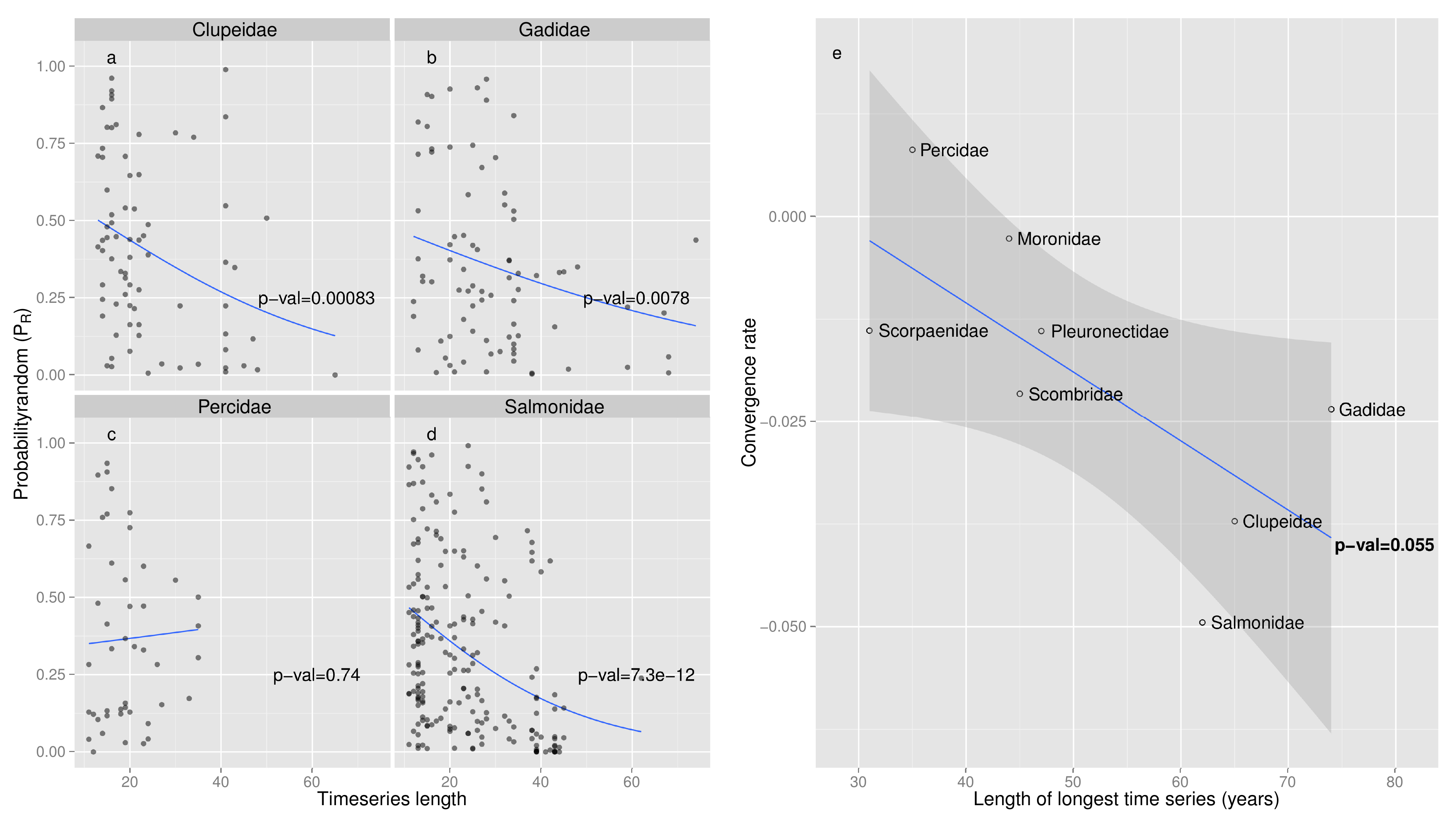}
\caption{{\bf Convergence across fish families. Probability random vs.\ time series length for families with at least 30 populations.} All families except Percidae exhibit convergence towards low probability random. Magnitude of the convergence rate (i.e., the slope of the beta-regression line in a-d) is proportional to the length of the longest time series for a family (e). Each point in (e) represents a single family in the RAM database (all families with at least 10 populations are shown); the blue line is the linear regression, the shaded area is the 95\% confidence interval.}
\end{figure}

%
%
%
\clearpage
\section*{Supporting Information}
%
%
\begin{figure}[htb]
\includegraphics[width=\textwidth]{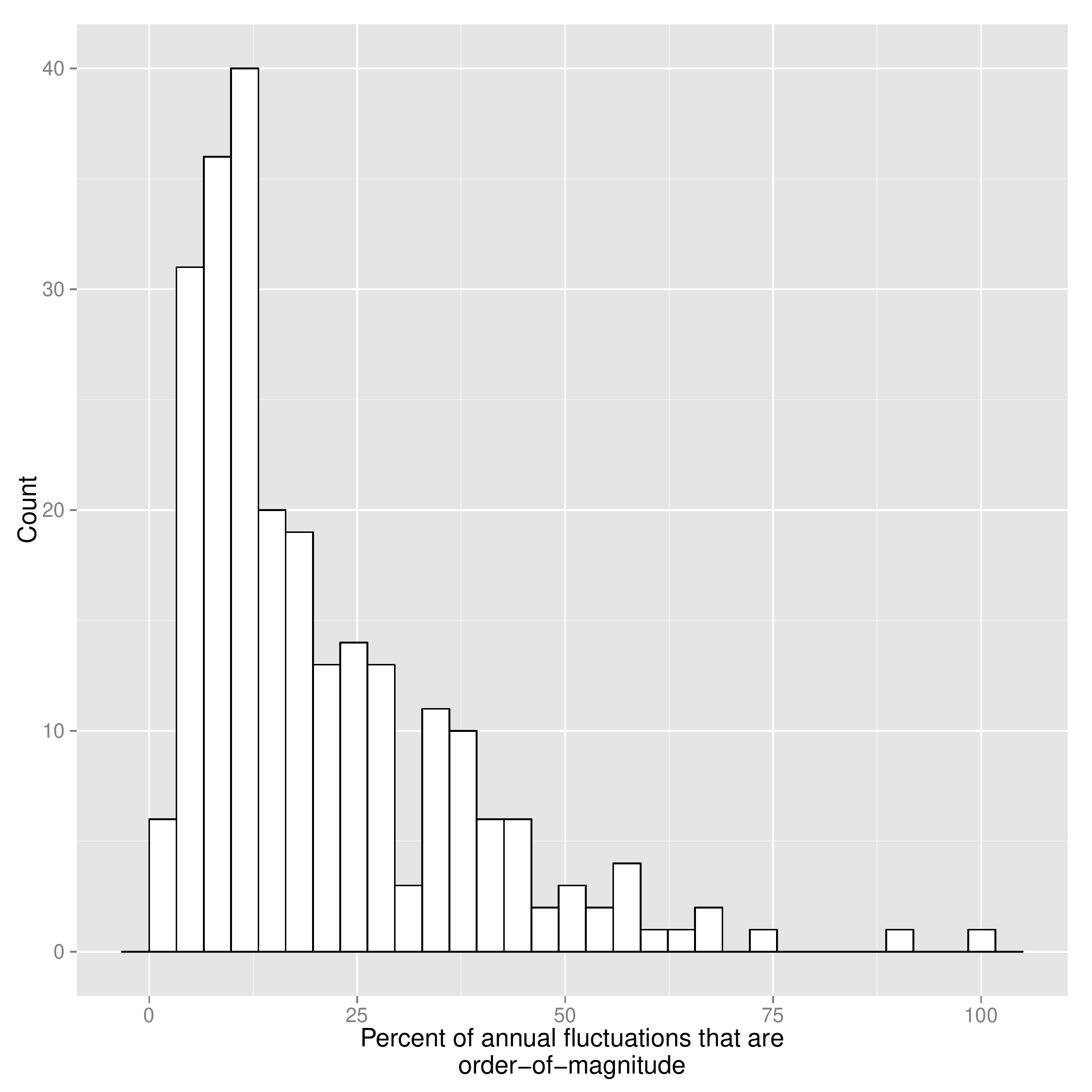}
\caption{Figure S1. Histogram of the percent of order-of-magnitude fluctuations in each population with at least one order-of-magnitude fluctuation. Overall, 36\% of the populations in the RAM database exhibited at least one order-of-magnitude fluctuation. Of those exhibiting at least one order-of-magnitude fluctuation, an order-of-magnitude fluctuation occurred approximately once every five years on average (i.e., 20\% of annual fluctuations were order-of-magnitude, on average).}
\end{figure}

\begin{figure}[htb]
\includegraphics[width=\textwidth]{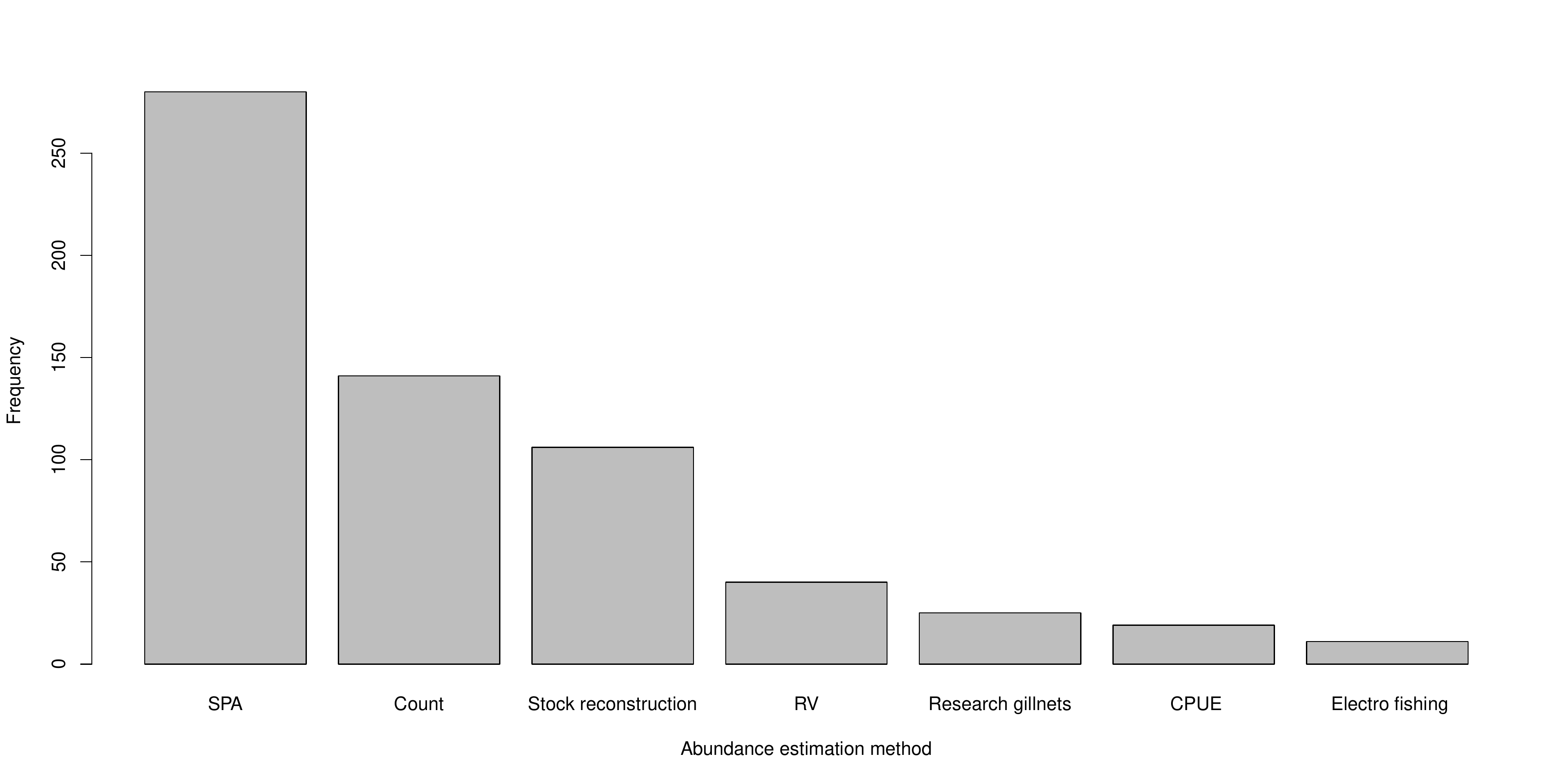}
\caption{Figure S2. Abundance estimation methods in the RAM database with at least 10 populations.}
\end{figure}

\begin{figure}[htb]
\includegraphics[width=\textwidth]{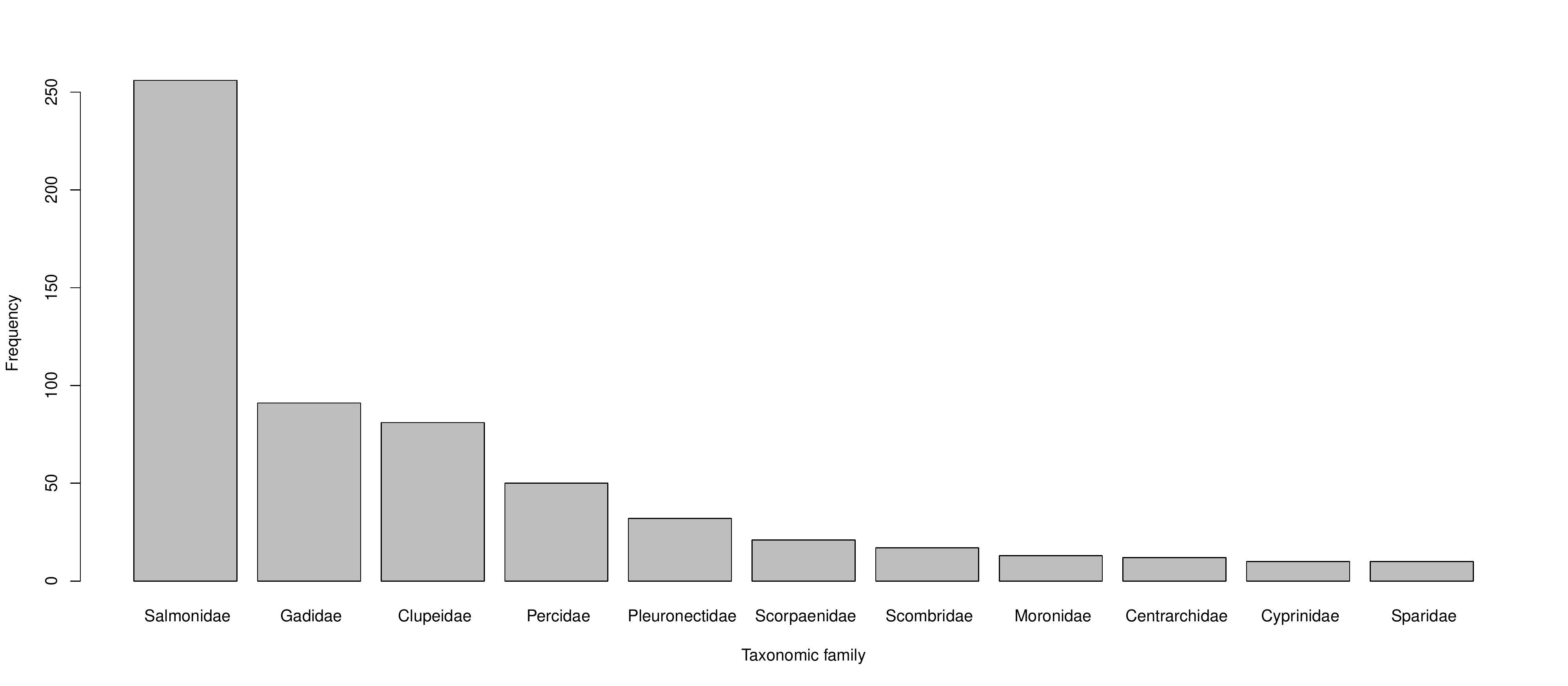}
\caption{Figure S3. Fish families in the RAM database with at least 10 populations.}
\end{figure}

\begin{figure}[htb]
\includegraphics[width=\textwidth]{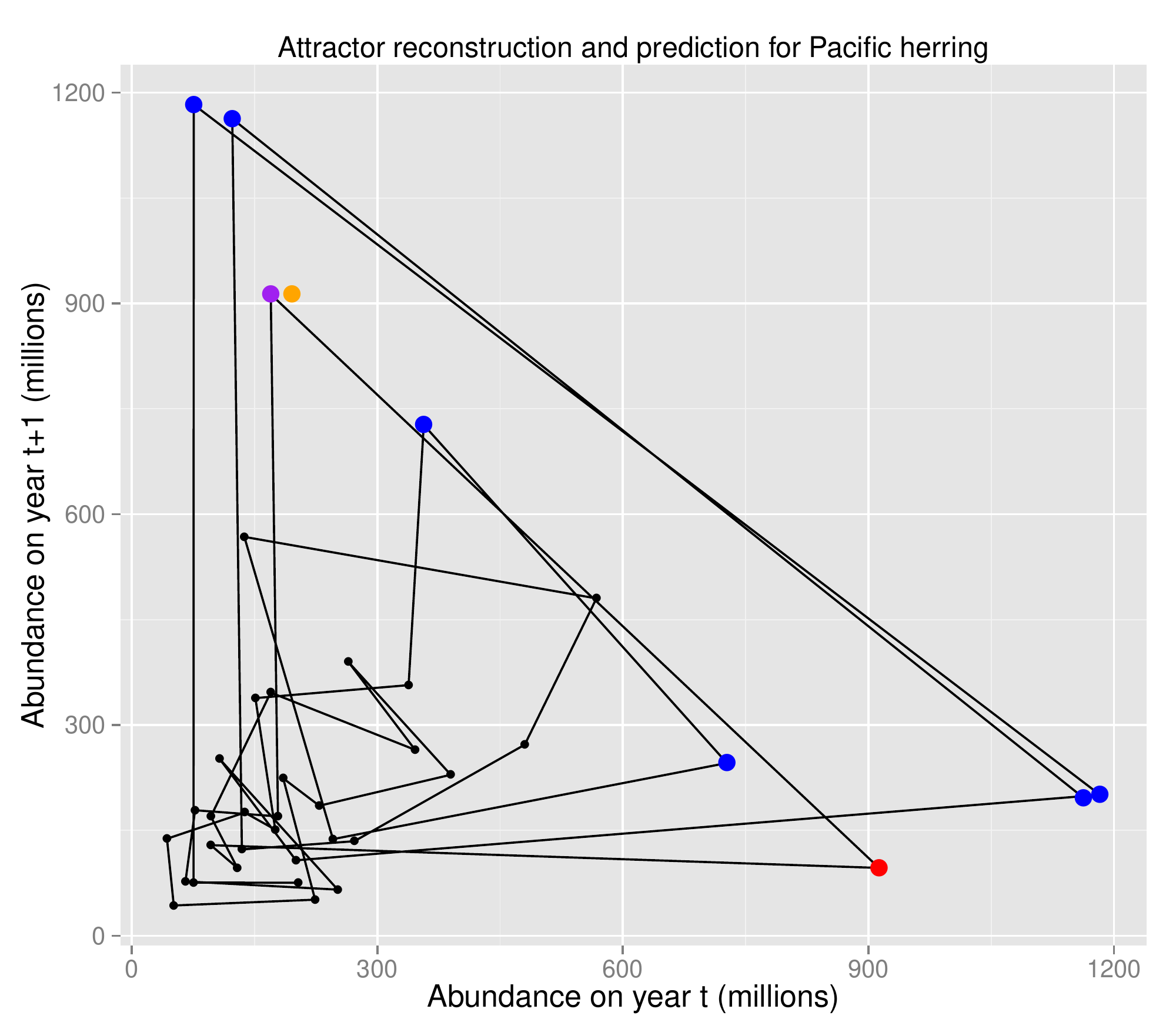}
\caption{Figure S4. Attractor reconstruction and example Simplex projection prediction for central coast British Columbia Pacific herring (\textit{Clupea pallasii}).The three points highlighted in blue closest to the red point are the \textit{E}+1 nearest neighbours which are used to predict the red point. The point highlighted in purple is the observed abundance, the orange point is the predicted abundance.}
\end{figure}

\begin{figure}[htb]
\includegraphics[width=\textwidth]{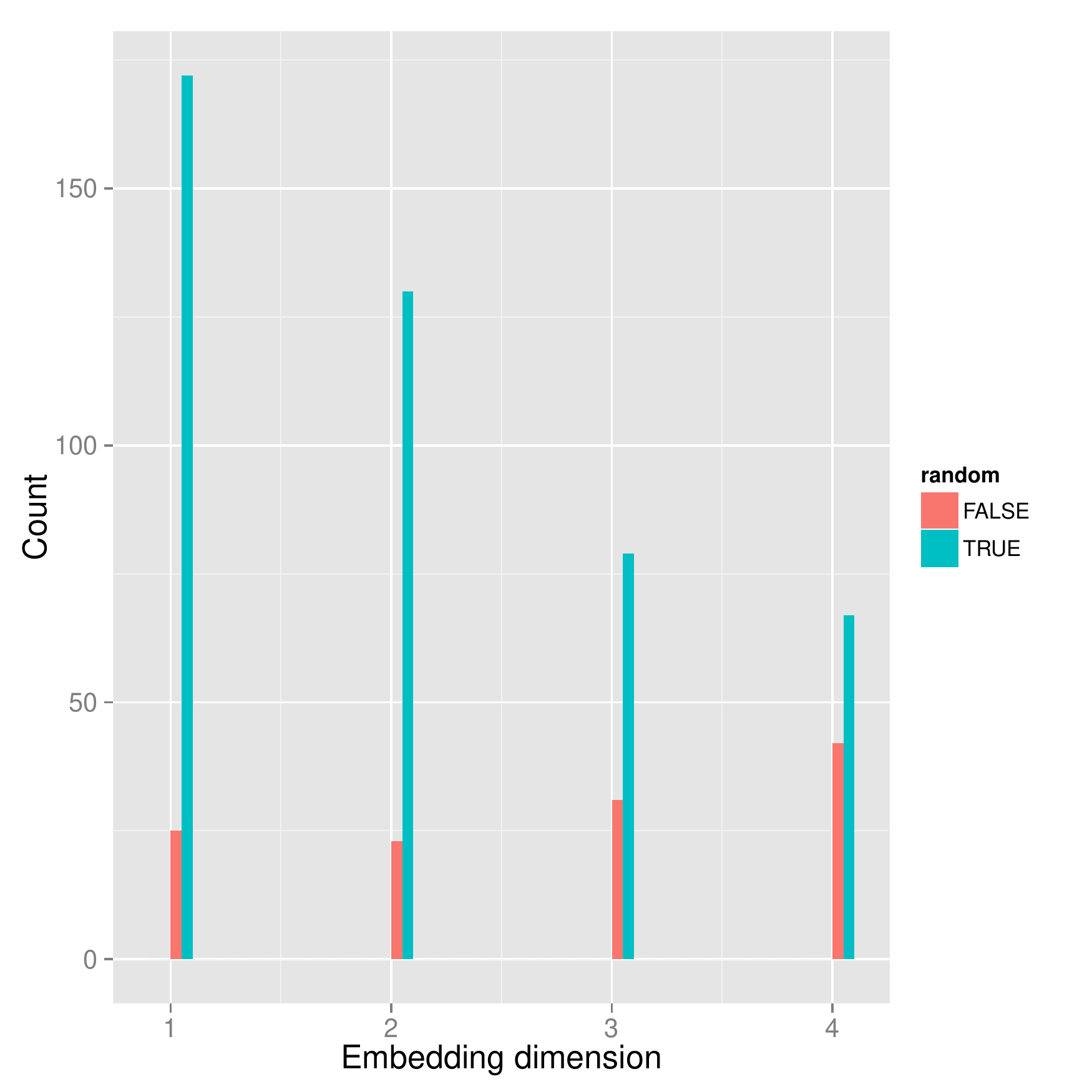}
\caption{Figure S5. Best-fit embedding dimension for all populations. Random populations are those that have a probability random greater than 0.1. The average embedding dimension is higher for the populations governed by non-random dynamics than those governed by random dynamics.}
\end{figure}

\begin{figure}[htb]
\includegraphics[width=\textwidth]{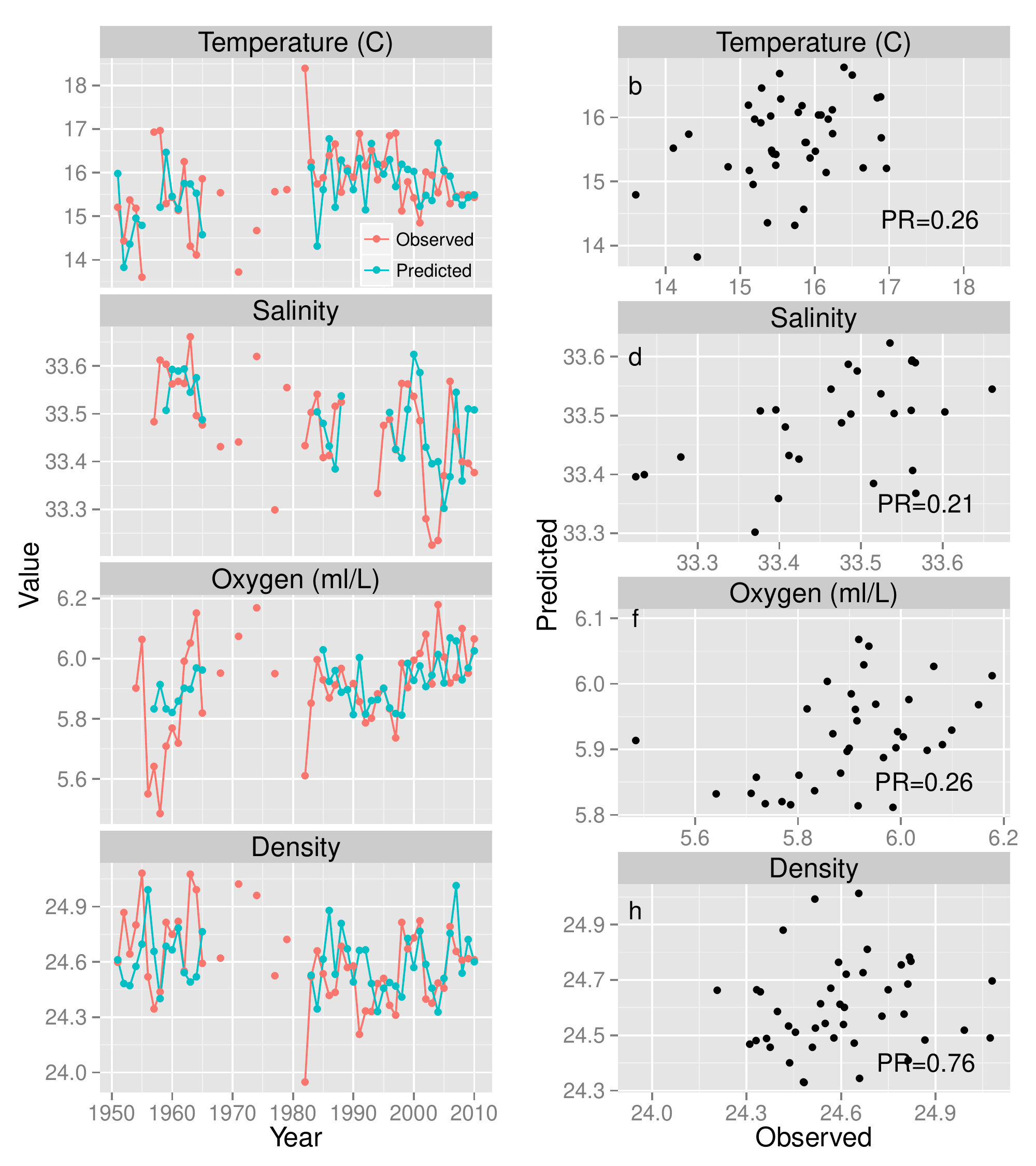}
\caption{ Figure S6. Environmental time series from the CalCOFI dataset (California Cooperative Oceanic Fisheries Investigations). Values are the annual spatial average of surface measurements in the core grid (lines 76.7 to 93.3).  Predicted and observed (left column), and predicted vs.\ observed (right column) are shown. None of the environmental time series are classified as non-random.}
\end{figure}

\begin{figure}[htb]
\includegraphics[width=\textwidth]{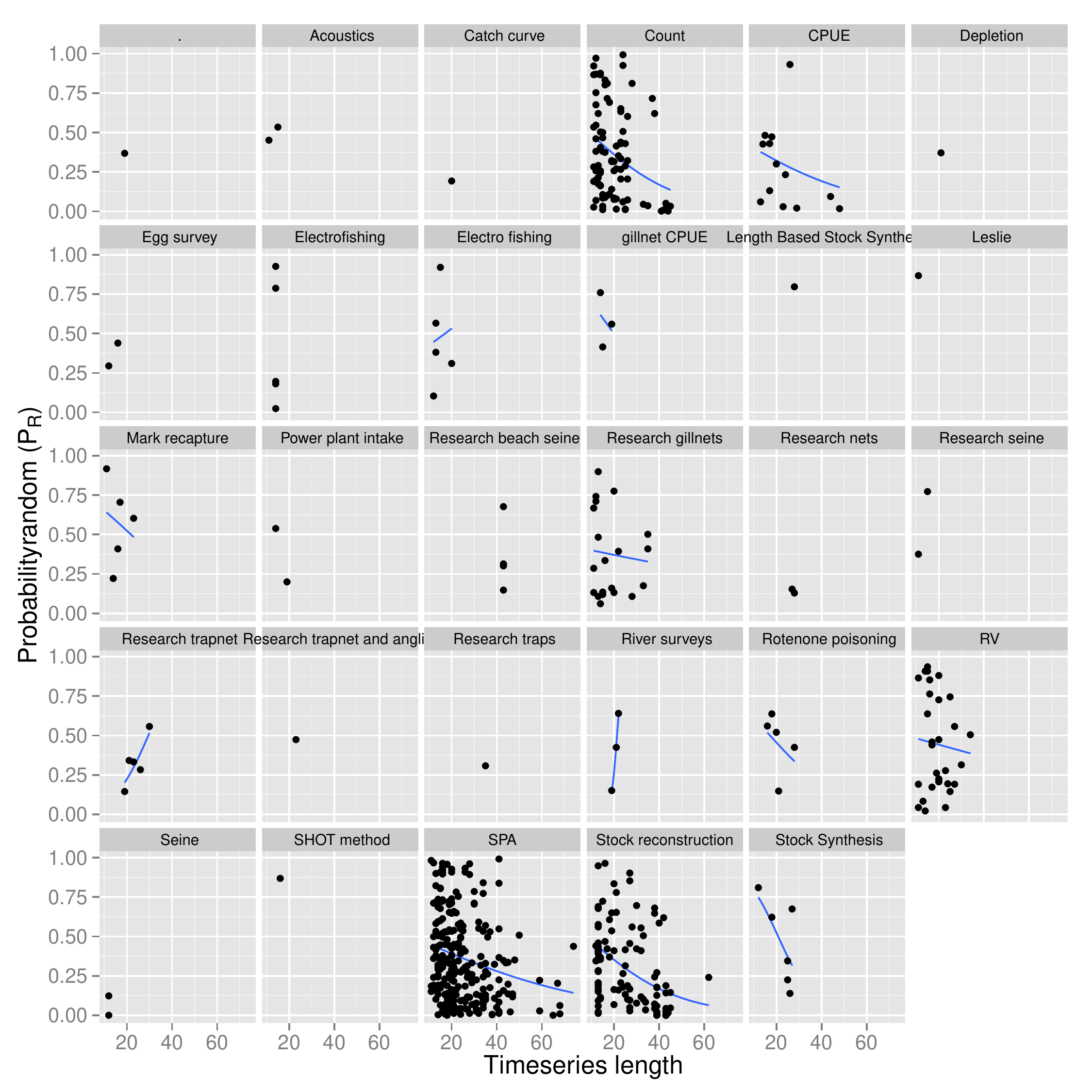}
\caption{ Figure S6. Convergence towards determinism across abundance estimation methods. Most methods exhibit a declining probability random with increasing time series length. A beta-regression was performed on all methods with at least three populations.}
\end{figure}

\begin{figure}[htb]
\includegraphics[width=\textwidth]{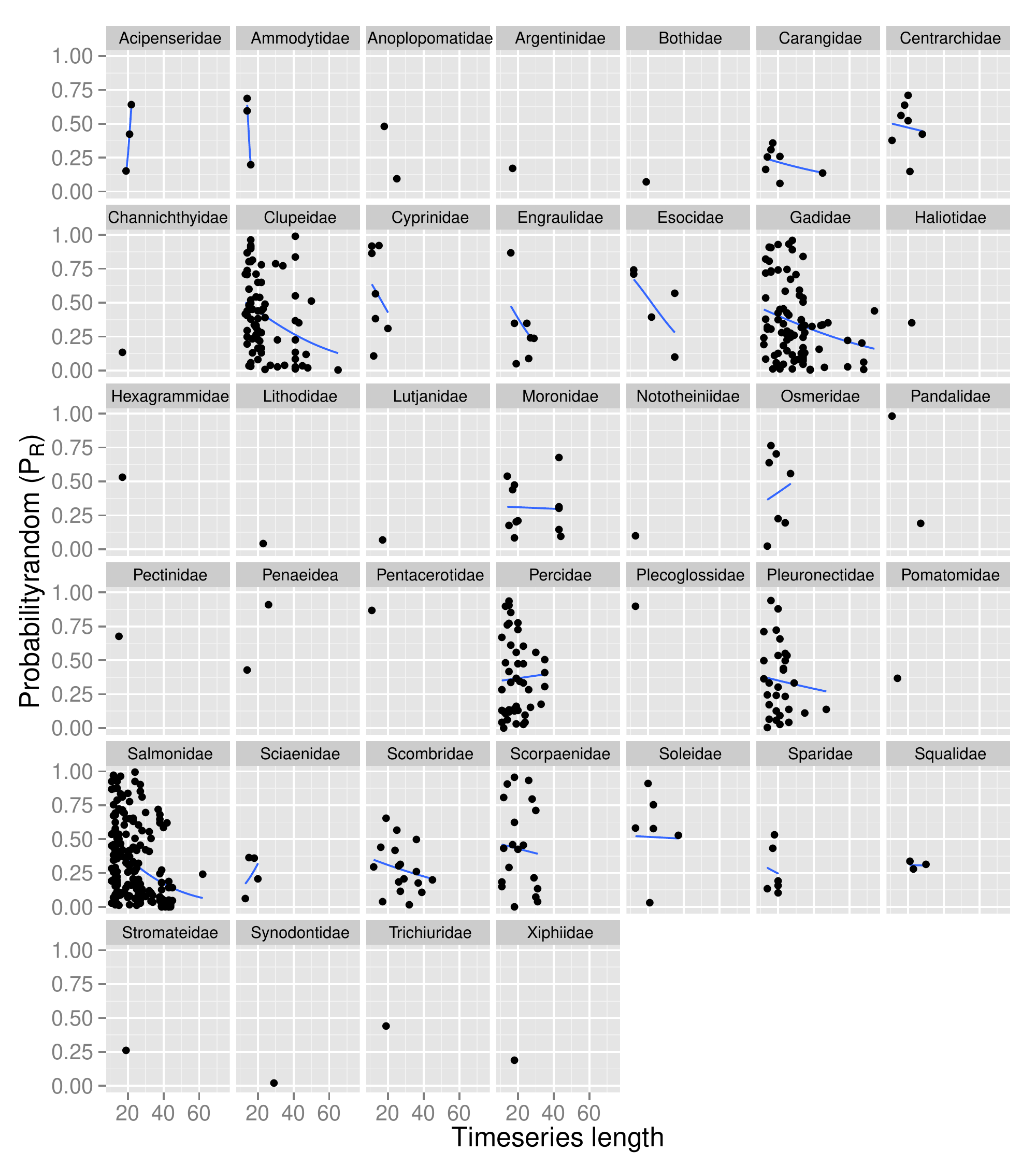}
\caption{Figure S7. Convergence towards non-random dynamics across families. Most families exhibit declining probability random with increasing time series length. A beta-regression was performed on all families with at least three populations.}
\end{figure}

\begin{figure}[htb]
\includegraphics[width=\textwidth]{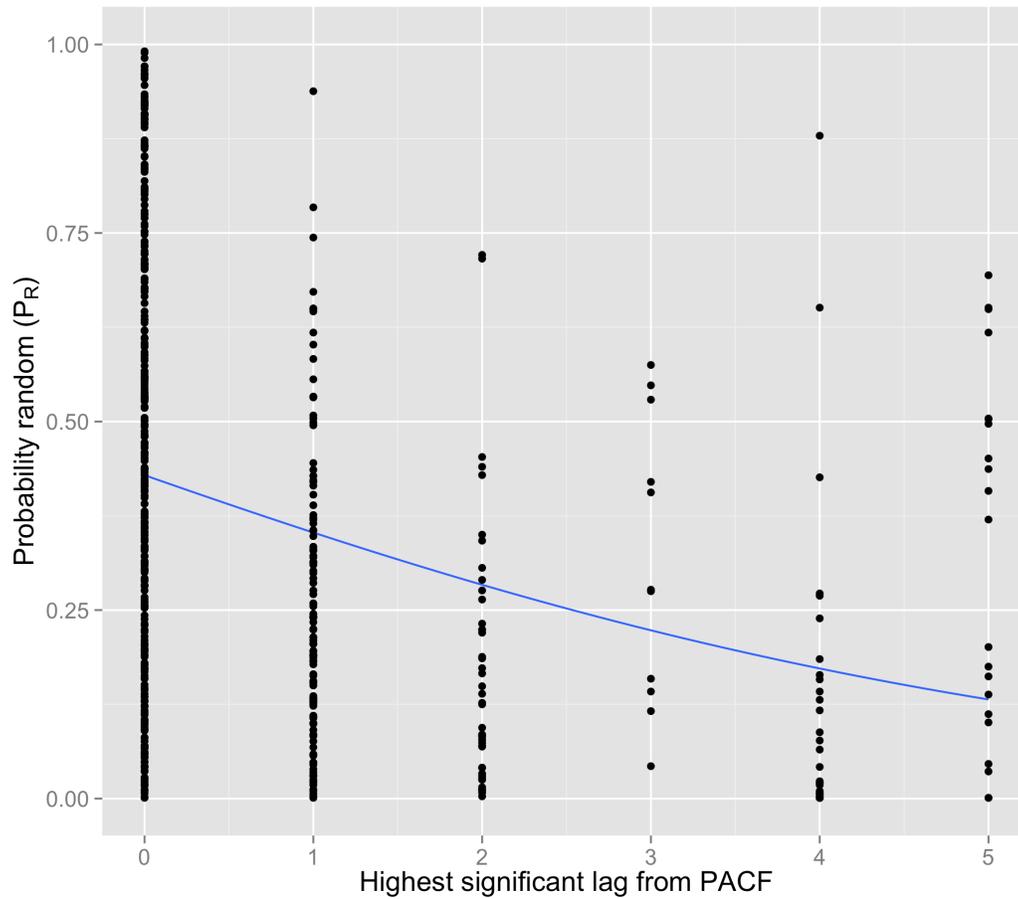}
\caption{Figure S8. Relationship between the highest significant lag from the partial autocorrelation function (PACF) and probability random. The PACF with a maximum lag of five was calculated for each population. The blue line represents the binomial regression of the highest significant lag from the PACF against the probability random calculated using the Monte Carlo procedure. Populations with low probability random were significantly more likely to exhibit significant higher-order lags in the PACF analysis ($p<0.001$) which provides supporting evidence that a non-random signal exists in recruitment time series.}
\end{figure}

\end{document}